# Generating Markov Equivalent Maximal Ancestral Graphs by Single Edge Replacement

Jin Tian
Department of Computer Science
Iowa State University
Ames, IA 50011
*jtian@cs.iastate.edu*

## Abstract

Maximal ancestral graphs (MAGs) are used to encode conditional independence relations in DAG models with hidden variables. Different MAGs may represent the same set of conditional independences and are called Markov equivalent. This paper considers MAGs without undirected edges and shows conditions under which an arrow in a MAG can be reversed or interchanged with a bi-directed edge so as to yield a Markov equivalent MAG.

## 1 Introduction

The use of graphical models for encoding distributional and causal assumptions is now fairly standard (see, for example, [Pearl, 1988, Spirtes et al., 1993, Pearl, 2000]). The most common such representation involves a directed acyclic graph (DAG), called a Bayesian network, over a set of variables. The statistical information encoded in a Bayesian network is completely captured by conditional independence relationships among the variables.

When some variables in a DAG model are not observed, called *latent* or *hidden* variables, the conditional independence relations among observed variables cannot, in general, be represented by a DAG with only the observed variables. In this paper we consider a class of graphs, called maximal ancestral graphs (MAGs) [Richardson and Spirtes, 2002], that capture the independence relations among observables without explicitly including latent variables in the model. MAGs contain three types of edges, directed edges ($\rightarrow$), bi-directed edges ($\leftrightarrow$), and undirected edges (—). Bi-directed edges are used to represent latent variables, and undirected edges are used to represent selection variables (variables that are conditioned upon).

Two different DAGs may represent the same set of conditional independence relations [Verma and Pearl, 1990] and are called *Markov equivalent*. Similarly, two different MAGs may also be Markov equivalent. A graphical characterization of Markov equivalence classes plays an important role in the applications of graphical models. We can not distinguish Markov equivalent graphs by knowledge about independence relations alone. In the process of learning graphical structures from data, it may have advantage to search over Markov equivalence classes instead of the individual graphs [Chickering, 2002a, Chickering, 2002b]. To interpret a learned model it is essential to know the features common to all graphs in an equivalence class.

The characterization of Markov equivalence classes for DAGs were given in [Verma and Pearl, 1990, Meek, 1995, Andersson et al., 1997]. A full characterization for equivalence classes of MAGs are still unavailable. [Ali and Richardson, 2002] sought to characterize those edges that are common to every graph in an equivalence class of MAGs. Graphical conditions for two MAGs being Markov equivalent were given in [Spirtes and Richardson, 1996, Ali et al., 2004].

A transformational characterization of Markov equivalent DAGs was given in [Chickering, 1995], which gave rules for transforming a DAG into a Markov equivalent DAG by edge reversing (replace $X \rightarrow Y$ by $X \leftarrow Y$) and showed that, for any two Markov equivalent DAGs, there exists a sequence of such arrow reversals that takes one DAG into the other. In this paper we seek a transformational characterization of Markov equivalence for MAGs without undirected edges.[1] We provide conditions under which a directed edge can be reversed, or interchanged with a bi-directed edge, to generate a Markov equivalent MAG. However, we were unable to decide whether any two Markov equiva-lent MAGs can be transformed into each other through such single edge replacement.

---
[1] Two Markov equivalent MAGs containing undirected edges may not be transformed to each other by single edge replacement (see Section 2).

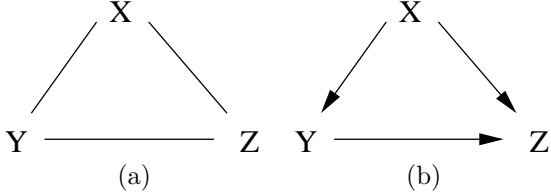

Figure 1: MAG (a) cannot be transformed into MAG (b) by single edge replacement.

Chickering's transformational characterization for DAGs has had important applications. We expect that a transformational characterization of Markov equivalent MAGs will have similar applications in DAG models with latent variables. For example, it might be useful for identifying those edges which are present in every MAG in an equivalence class; it may facilitate a score-based structure learning algorithm that searches on the space of Markov equivalence classes of MAGs.

The rest of the paper is organized as follows. Section 2 clarifies the reason why we only consider MAGs without undirected edges. Section 3 introduces MAGs and provides basic notation, definitions, and theorems. Section 4 contains the main result of the paper, with all the proofs given in the Appendix.

## 2 Undirected Edges in a MAG

In a general MAG, undirected edges are used to represent the presence of selection variables, that is, variables that are conditioned upon. The definition of a MAG leads to that the configurations $X - Y \leftrightarrow Z$ and $X - Y \leftarrow Z$ do not occur, and therefore any MAG can be split into an undirected graph $G_u$ and a MAG containing no undirected edges $G_a$ such that any edge between a node $X$ in $G_u$ and a node $Y$ in $G_a$ takes the form $X \to Y$ [Richardson and Spirtes, 2002]. In general, two Markov equivalent MAGs containing undirected edges may not be transformed to each other by single edge replacement. For example, the two MAGs in Figure 1 are Markov equivalent, however any single replacement of an undirected edge by a directed edge in Figure 1(a) will lead to a graph that is not a MAG. For this reason, in this paper, we will only consider MAGs without undirected edges. The results will be useful in models containing latent variables but no selection variables.

## 3 Notation, Definitions, and Theorems

We have mostly adopted the notation and definitions in [Ali and Richardson, 2002]. If there is an edge between $X$ and $Y$, then $X$ is *adjacent* to $Y$. If $X \to Y$, then $X$ is a *parent* of $Y$ and $Y$ is a *child* of $X$. If $X \leftrightarrow Y$, then $X$ is a *spouse* of $Y$. A *directed path* from $X$ to $Y$ is a path composed of directed edges ($X \to \ldots \to Y$). If there is a directed path from $X$ to $Y$ or $X = Y$, then $X$ is an *ancestor* of $Y$. A node $Z$ on a path is called a *collider* if two arrowheads on the path meet at $Z$, i.e. $X \to Z \leftarrow Y, X \leftrightarrow Z \leftrightarrow Y$, $X \leftrightarrow Z \leftarrow Y$, or $X \to Z \leftrightarrow Y$; all other non-endpoint nodes on a path are called *non-colliders*, i.e. $X \leftarrow Z \to Y, X \leftarrow Z \leftarrow Y, X \to Z \to Y, X \leftrightarrow Z \to Y$, or $X \leftarrow Z \leftrightarrow Y$. A collider (non-collider) is said to be *unshielded* if $X$ and $Y$ above are not adjacent, and *shielded* otherwise.

Ancestral graphs have at most one edge between each pair of nodes, and in this paper we consider ancestral graphs without undirected edges.

**Definition 1 (ancestral graph)** *A graph which may contain directed ($\to$) or bi-directed edges ($\leftrightarrow$) is* ancestral *if*

1. *there are no directed cycles;*

2. *whenever there is an edge $X \leftrightarrow Y$, then there is no directed path from $X$ to $Y$, or from $Y$ to $X$.*

Pearl's d-separation criterion can be naturally extended to be applied in ancestral graphs, and are called m-separation whose formal definition is given below.

**Definition 2 (m-separation)** *A path between nodes $X$ and $Y$ in an ancestral graph is said to be* m-connecting given a set $Z$ if

1. *every non-collider on the path is not in $Z$, and*

2. *every collider on the path is an ancestor of a node in $Z$.*

*If there is no path m-connecting $X$ and $Y$ given $Z$, then $X$ and $Y$ are said to be* m-separated *given $Z$. Two Sets $V$ and $W$ are said to be* m-separated *given $Z$, if for every pair $X, Y$, with $X \in V$ and $Y \in W$, $X$ and $Y$ are m-separated given $Z$.*

M-separation characterizes the independence relations represented by an ancestral graph.

**Definition 3 (Markov equivalent)** *Two graphs $G_1$ and $G_2$ are said to be Markov equivalent if for all disjoint sets $A, B, Z$, $A$ and $B$ are m-separated given $Z$* in $G_1$ if and only if $A$ and $B$ are m-separated given $Z$ in $G_2$.

In a DAG, two nodes are m-separable by some set of variables if and only if they are not adjacent. It is no longer the case in an ancestral graph. It is possible that two non-adjacent nodes are not m-separable by any set of variables. Since we are only concerned with independence relations, we may add an edge between non-adjacent but unseparable nodes, which motivates the following definition.

**Definition 4 (maximal ancestral graph (MAG))** *An ancestral graph is said to be maximal if, for every pair of non-adjacent nodes $X, Y$ there exists a set $Z$ such that $X$ and $Y$ are m-separated conditional on $Z$.*

Every non-maximal ancestral graph corresponds to a MAG, so we will restrict our attention to MAGs in the rest of the paper.

The conditions for unseparability are characterized by the notion of *inducing path* [Verma and Pearl, 1990].

**Definition 5 (inducing path)** *A path between two nodes $X$ and $Y$ is an inducing path if and only if every internal node in the path is a collider on the path and is an ancestor of $X$ or $Y$ or both.*

The following theorem gives conditions for an arbitrary graph (singly connected with bidirected edges) to be a MAG.

**Theorem 1** *[Spirtes et al., 1997] A graph $M$ is a MAG if and only if:*

1. *If there is an inducing path between $X$ and $Y$ in $M$, then $X$ and $Y$ are adjacent in $M$.*

2. *If there is an edge $X \longrightarrow Y$ in $M$, then $Y$ is not an ancestor of $X$ in $M$.*

3. *If there is an edge $X \longleftrightarrow Y$ in $M$, then $X$ is not an ancestor of $Y$ and $Y$ is not an ancestor of $X$ in $M$.*

The conditions for two MAGs to be Markov equivalent are characterized by the following definition and theorem.

**Definition 6 (discriminating path)** *In an ancestral graph $G$, a path $p$ between nodes $X$ and $Y$ is a discriminating path for node $Z$ if and only if*

1. *$p$ contains $Z$, $Z \neq X$, $Z \neq Y$,*

2. *$p$ has at least three edges,*

3. *$Z$ is adjacent to $Y$ on $p$, $X$ is not adjacent to $Y$, and*

4. *for every node $Q$ between $X$ and $Z$ on $p$ except for the endpoints, $Q$ is a collider on $p$ and there is an edge $Q \rightarrow Y$ in $G$.*

**Theorem 2** *[Spirtes and Richardson, 1996]* *Two MAGs $M_1$ and $M_2$ are Markov equivalent if and only if*

1. *they have the same adjacencies;*

2. *they have the same set of unshielded colliders; and*

3. *if $p$ is a discriminating path for $X$ in $M_1$ and the corresponding path $p'$ in $M_2$ is a discriminating path for $X$, then $X$ is a collider on $p$ in $M_1$ if and only if $X$ is a collider on $p'$ in $M_2$.*

[Ali *et al.*, 2004] gives another (stronger) criterion for two MAGs being Markov equivalent. However, our results in this paper are based on Theorem 2.

## 4 Generating Markov Equivalent MAGs

[Chickering, 1995] gave conditions for edge reversals in a DAG to generate a Markov equivalent DAG. In this section we give conditions for single edge replacement in a MAG to generate a Markov equivalent MAG based on Theorem 2. Proofs of theorems are given in the Appendix.

**Definition 7 (blanketed edge)** *An edge $X \longrightarrow Y$ is a* blanketed edge *in a MAG if*

1. *there is no directed path from $X$ to $Y$ (other than $X \longrightarrow Y$ itself),*

2. *all parents of $X$ are parents of $Y$, and*

3. *for each $X$'s spouse $Z$, either $Z$ is a spouse of $Y$, or $Z$ is a parent of $Y$ and there is no discriminating path for $X$ that contains the path $Z \longleftrightarrow X \longrightarrow Y$.*

*An edge $X \longleftrightarrow Y$ is a* blanketed edge against $X$ *in a MAG if the above conditions 2 and 3 are satisfied.*[2]

Note that whether an edge is a blanketed edge can be determined efficiently as for a fixed triplet $X, Y, Z$ in the above condition 3, whether there is a discriminating path can be determined in time linear in the number of edges in the MAG.

---
[2] Condition 1 is always satisfied in a MAG containing $X \longleftrightarrow Y$.

The following theorem gives conditions for exchanging a directed edge with a bi-directed edge in a MAG.

**Theorem 3** 1. Let $M$ be a MAG with the edge $X \longrightarrow Y$, and let $M'$ be a MAG with the edge $X \longleftrightarrow Y$ otherwise identical to $M$. If $M'$ is Markov equivalent to $M$, then $X \longrightarrow Y$ is a blanketed edge in $M$ and $X \longleftrightarrow Y$ is a blanketed edge against $X$ in $M'$.

2. If $X \longrightarrow Y$ is a blanketed edge in a MAG $M$, replacing it by $X \longleftrightarrow Y$ will result in a MAG and the MAG is Markov equivalent to $M$.

3. If $X \longleftrightarrow Y$ is a blanketed edge against $X$ in a MAG $M$, replacing it by $X \longrightarrow Y$ will result in a MAG and the MAG is Markov equivalent to $M$.

We use $Pa(X)$ to denote the set of parents of $X$ and $Sp(X)$ to denote the set of spouses of $X$.

**Definition 8 (screened edge)** An edge $X \longrightarrow Y$ is a screened edge in a MAG if $Pa(Y) = Pa(X) \cup \{X\}$ and $Sp(Y) = Sp(X)$.

The following theorem gives conditions for reversing a directed edge in a MAG.

**Theorem 4** Let $M$ be a MAG with the edge $X \longrightarrow Y$, and let $M'$ be a graph with the edge $X \longleftarrow Y$ otherwise identical to $M$. Then $M'$ is a MAG that is Markov equivalent to $M$ if and only if $X \longrightarrow Y$ is a screened edge in $M$.

[Chickering, 1995] showed that for two Markov equivalent DAGs there always exists a sequence of edge reversals that will transform one DAG into the other. In order to prove a similar property in MAGs, we need to show the following. For two MAGs $M$ and $M'$ with the same set of nodes and the same adjacencies, we use $\Delta(M, M')$ to represent the set of edges in $M$ that are different in $M'$.

**Conjecture 1** If two MAGs $M$ and $M'$ are Markov equivalent, then there exists an edge in $\Delta(M, M')$ that is blanketed or blanketed against some node.

If the conjecture holds, then $M$ can be transformed to $M'$ by a sequence of adding/removing arrowheads. Currently we are unable to prove or refute this conjecture.

With Theorems 3 and 4, given a MAG $M$, we can generate a set of MAGs that are Markov equivalent to $M$ by repeated single edge replacement. Whether all the equivalent MAGs can be generated this way depends on the trueness of Conjecture 1.

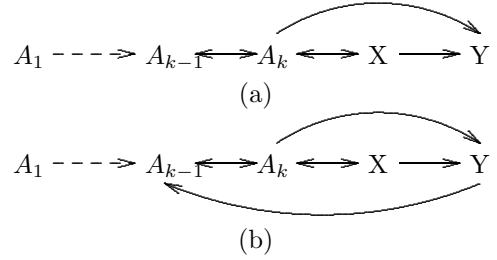

Figure 2: Lemma 1

## 5 Conclusion

We give conditions under which a single edge replacement in a MAG leads to a Markov equivalent MAG. The result can be used to generate equivalent MAGs starting from a given MAG. Whether all the equivalent MAGs can be generated this way remains an open problem. The result is one step toward the goal of a full transformational characterization of Markov equivalent MAGs.

### Acknowledgements

The author thanks the anonymous reviewers for helpful comments. This research was partly supported by NSF grant IIS-0347846.

## Appendix

A path $p$ between two nodes $X$ and $Y$ may be denoted by $p = (X, \ldots, Z, \ldots, Y)$ where $Z$ is a node on $p$. A subpath of $p$ between two nodes $W$ and $Z$ on $p$ will be denoted by $p(W, Z)$.

**Lemma 1** Let $X \longrightarrow Y$ (or $X \longleftrightarrow Y$) be a blanketed edge (against $X$) in a MAG $M$. If there is a path $p = (A_1, \ldots, A_k, X)$ between $A_1$ and $X$ such that every internal node is a collider on $p$ and $A_k$ is a spouse of $X$, then for $i = 1, \ldots, k$ either there exists a node $A_i$ being a spouse of $Y$ or every node $A_i$ is a parent of $Y$.

**Proof:** Assume none of $A_i$ is a spouse of $Y$. Since $A_k$ is a spouse of $X$, $A_k$ must be a parent of $Y$ from the definition of a blanketed edge. $A_{k-1}$ must be adjacent to $Y$ else the path $(A_{k-1}, A_k, X, Y)$ would be a discriminating path for $X$ (see Figure 2(a)). $A_{k-1}$ cannot be a child of $Y$ else $(A_k, Y, A_{k-1})$ is a directed path from $A_k$ to $A_{k-1}$ which contradicts to the fact that $M$ is a MAG (see Figure 2(b)). Thus if $A_{k-1}$ is not a spouse of $Y$ it must be a parent of $Y$. With the same arguments we can show $A_i$, $i = 1, \ldots, k$, must all be parents of $Y$. □

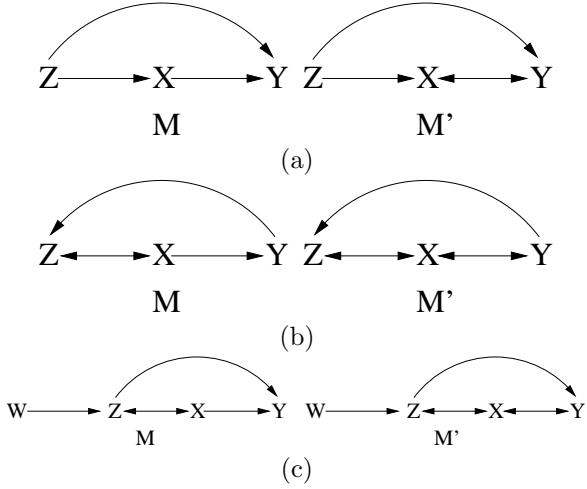

Figure 3: Theorem 3-1

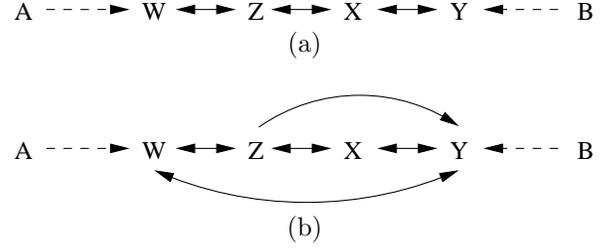

Figure 4: Theorem 3-2

**Theorem 3**  1. Let $M$ be a MAG with the edge $X \longrightarrow Y$, and let $M'$ be a MAG with the edge $X \longleftrightarrow Y$ otherwise identical to $M$. If $M'$ is Markov equivalent to $M$, then $X \longrightarrow Y$ is a blanketed edge in $M$ and $X \longleftrightarrow Y$ is a blanketed edge against $X$ in $M'$.

2. If $X \longrightarrow Y$ is a blanketed edge in a MAG $M$, replacing it by $X \longleftrightarrow Y$ will result in a MAG and the MAG is Markov equivalent to $M$.

3. If $X \longleftrightarrow Y$ is a blanketed edge against $X$ in a MAG $M$, replacing it by $X \longrightarrow Y$ will result in a MAG and the MAG is Markov equivalent to $M$.

**Proof:**

1. Since $M'$ contains the edge $X \longleftrightarrow Y$, from the definition of a MAG there is no directed path from $X$ to $Y$. If $Z$ is a parent of $X$, then $Z$ must be adjacent to $Y$, else there will be a unshielded collider in $M'$ but not in $M$ (see Figure 3(a)). From the definition of a MAG, the edge between $Z$ and $Y$ can only be $Z \longrightarrow Y$. If $Z$ is a spouse of $X$ then $Z$ must be adjacent to $Y$ else there will be a unshielded collider in $M'$ which is not in $M$. $Z$ cannot be a child of $Y$ else $(X, Y, Z)$ is a directed path from $X$ to $Z$ in $M$ which contradicts the fact that $M$ is a MAG (see Figure 3(b)). $Z$ can be a spouse of $Y$. If $Z$ is a parent of $Y$, then there is no discriminating path for $X$ that contains the path $Z \longleftrightarrow X \longrightarrow Y$ in $M$, else the same path would be a discriminating path for $X$ in $M'$ which contradicts the fact that $M$ and $M'$ are Markov equivalent (see Figure 3(c)).

2. Let $M'$ be the graph resulted from replacing $X \longrightarrow Y$ by $X \longleftrightarrow Y$ in $M$. First we show that $M'$ is a MAG. First of all, since $X \longrightarrow Y$ is a blanketed edge there is no directed path from $X$ to $Y$ or from $Y$ to $X$. Replacing $X \longrightarrow Y$ by $X \longleftrightarrow Y$ will not create a new directed path. We need to show that the edge $X \longleftrightarrow Y$ will not create an inducing path (with non-adjacent endpoints). Assume that replacing $X \longrightarrow Y$ by $X \longleftrightarrow Y$ leads to an inducing path $p$ between two nodes $A$ and $B$. $p$ must contain $X \longleftrightarrow Y$ in $M'$ else $p$ would already be an inducing path in $M$ (see Figure 4(a)). Each internal node on $p$ between $A$ and $X$ is a collider, thus by Lemma 1 either there exists a node being a spouse of $Y$ or every node from $A$ to $Z$ is a parent of $Y$. If a node $W$ is a spouse of $Y$, then the path $(p(A, W), W, Y, p(Y, B))$ will already be an inducing path in $M$(see Figure 4(b)). If none of the internal node between $A$ and $X$ is a spouse of $Y$, then $A$ is a parent or spouse of $Y$, but then the path $(A, Y, p(Y, B))$ will already be an inducing path in $M$.

Next we show that $M'$ is Markov equivalent to $M$. First, $M'$ has the same adjacencies as $M$. Second, if $M'$ has an unshielded collider that is not in $M$, the unshielded collider must involve the edge $X \longleftrightarrow Y$ and $X$'s parents or spouses. However all $X$'s parents and spouses are adjacent to $Y$. Finally if $Z$ is a parent or spouse of $X$ there is no discriminating path for $X$ that contains the path $(Z, X, Y)$ in $M'$ from the definition of a blanketed edge.

3. If $X \longleftrightarrow Y$ is a blanketed edge against $X$ in a MAG $M$, let $M'$ be the graph with the edge $X \longrightarrow Y$ otherwise identical to $M$. First we show that $M'$ is a MAG. $M'$ cannot have a cycle since there is no directed path from $Y$ to $X$ in $M$ from the definition of a MAG. The operation of changing edge $X \longleftrightarrow Y$ to $X \longrightarrow Y$ may make $M'$ not a MAG in two situations: (1) the operation creates a directed path between two nodes connected by a bi-directed edge (Figure 5(a)); (2) the operation creates an inducing path between two nonadjacent nodes (Figure 5(b)). In both Figure 5 (a) and (b) $A$ is not an ancestor of $B$ in $M$ but will be an ancestor of $B$ in $M'$. If $A$ does not coincide with $X$, then since a parent of $X$ is a parent of $Y$ from the definition of a blanketed edge, $A$ will already be an ancestor of $B$ in $M$. Thus we only need to consider the cases of $X = A$ which are shown in Figure 5(c) and (d). In Figure 5(c), $B$ is a spouse of $X$. From the definition of a blanketed edge $B$ either is a spouse of $Y$ or a parent of $Y$. However, since there is a directed path from $Y$ to $B$, from the definition of a MAG $B$ can neither be a parent nor a spouse of $Y$. Thus this case cannot happen. In Figure 5(d), let $p$ denote the inducing path between $C$ and $B$. Each node between $X$ and $B$ is a collider on $p$, thus by Lemma 1 either one node between $X$ and $B$ (including $B$) is a spouse of $Y$ or all of them are parents of $Y$. However $B$ can neither be a parent nor a spouse of $Y$ since there is a directed path from $Y$ to $B$. Therefore one internal node on path $p(X, B)$ must be a spouse of $Y$, say node $E$. Again by Lemma 1, either one internal node (say $F$) on path $p(X, C)$ is a spouse of $Y$ (see Figure 5(e)) or $C$ is a parent or spouse of $Y$ (see Figure 5(f)). In Figure 5(e) the path $(p(C, F), F, Y, E, p(E, B))$ will be an inducing path in $M$. In Figure 5(f) the path $(C, Y, E, p(E, B))$ will be an inducing path in $M$.

Next we show that $M'$ is Markov equivalent to $M$. First, $M'$ has the same adjacencies as $M$. Second, if $M$ has an unshielded collider not in $M'$, the unshielded collider must involve the edge $X \longleftrightarrow Y$ and $X$'s parents or spouses. However all $X$'s parents and spouses are adjacent to $Y$. Finally for all $Z$ which is a parent or spouse of $X$ there is no discriminating path for $X$ in $M$ that contains the path $(Z, X, Y)$ from the definition of a blanketed edge.

□

**Lemma 2** *If $X \longrightarrow Y$ is a screened edge in a MAG $M$, then it is a blanketed edge in $M$.*

**Proof:** We only need to show that there is no directed path from $X$ to $Y$. If there is a directed path from $X$ to $Y$ in $M$, let $Z$ be the node nearest to $Y$ in the path (see Figure 6). $Z$ is a parent of $Y$ thus $Z$ must be a parent of $X$ from the definition of a screened edge. Then there would be a cycle in $M$ (see Figure 6). □

**Theorem 4** *Let $M$ be a MAG with the edge $X \longrightarrow Y$, and let $M'$ be a graph with the edge $X \longleftarrow Y$ otherwise identical to $M$. Then $M'$ is a MAG that is Markov equivalent to $M$ if and only if $X \longrightarrow Y$ is a screened edge in $M$.*

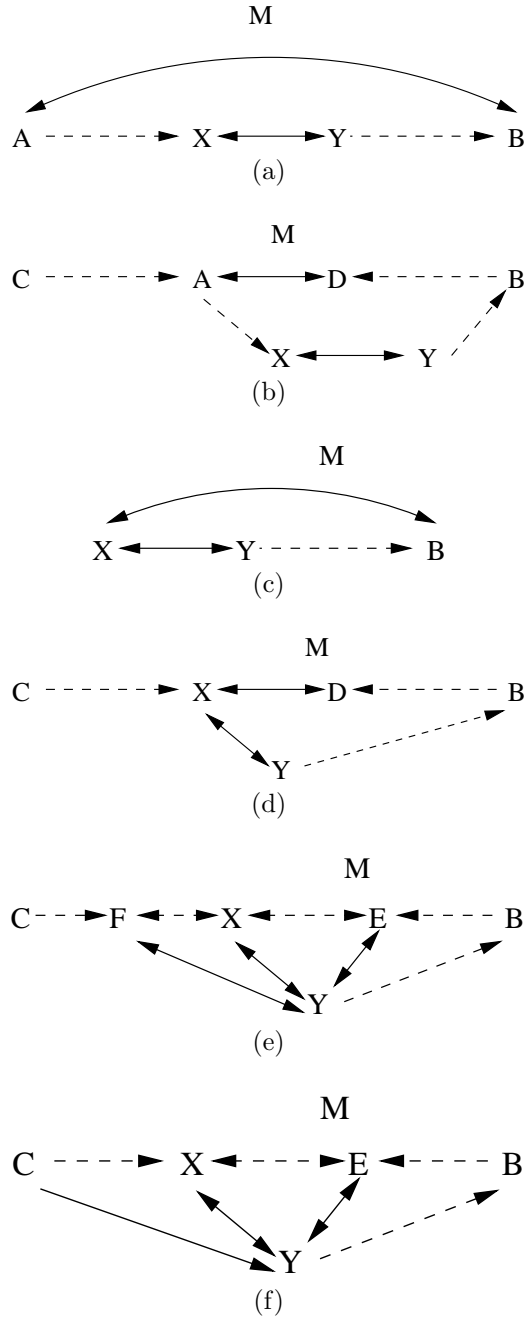

Figure 5: Theorem 3-3

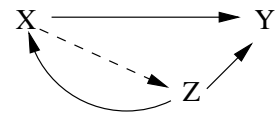

Figure 6: Lemma 2

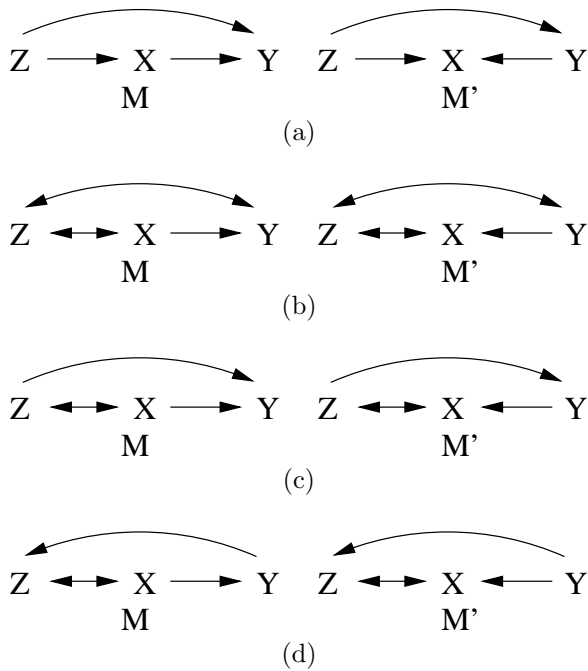

Figure 7: Theorem 4

**Proof:**
**(only if)** Let $Z$ be a parent of $X$. $Z$ must be adjacent to $Y$ else there is a unshielded collider in $M'$ which is not in $M$ (see Figure 7(a)). The edge between $Z$ and $Y$ must be $Z \longrightarrow Y$ because $M$ is a MAG. With similar arguments, a parent of $Y$ must be a parent of $X$.

Let $Z$ be a spouse of $X$. $Z$ must be adjacent to $Y$ else there is a unshielded collider in $M'$ which is not in $M$ (see Figure 7(b)). If the edge between $Z$ and $Y$ is $Z \longrightarrow Y$ (see Figure 7(c)), there will be a directed path from $Z$ to $X$ in $M'$ and $M'$ will not be a MAG. If the edge between $Z$ and $Y$ is $Z \longleftarrow Y$ (see Figure 7(d)), there will be a directed path from $X$ to $Z$ in $M$ and $M$ will not be a MAG. Thus $Z$ must be a spouse of $Y$. With similar arguments, a spouse of $Y$ must be a spouse of $X$.

**(if)** By Lemma 2, the edge $X \longrightarrow Y$ is a blanketed edge. Let $M''$ be the graph resulted from replacing $X \longrightarrow Y$ with $X \longleftrightarrow Y$ in $M$. Then $M''$ is a MAG Markov equivalent to $M$ by Theorem 3. $X \longleftrightarrow Y$ is a blanketed edge against $Y$ in $M''$ from the definition of a screened edge. Thus $M'$ is a MAG Markov equivalent to $M''$ by Theorem 3. Therefore $M'$ is a MAG Markov equivalent to $M$. □

# References

[Ali and Richardson, 2002] Ayesha Ali and Thomas Richardson. Markov equivalence classes for maximal ancestral graphs. In *Proceedings of the 18th Annual Conference on Uncertainty in Artificial Intelligence (UAI-02)*, pages 1–9, San Francisco, CA, 2002. Morgan Kaufmann Publishers.

[Ali *et al.*, 2004] Ayesha Ali, Thomas Richardson, and Peter Spirtes. Markov equivalence for ancestral graphs. Technical Report 465, University of Washington, Department of Statistics, September 2004.

[Andersson *et al.*, 1997] S.A. Andersson, D. Madigan, and M.D. Perlman. A characterization of Markov equivalence classes for acyclic digraphs. *Annals of Statistics*, 25:505–541, 1997.

[Chickering, 1995] D.M. Chickering. A transformational characterization of Bayesian network structures. In P. Besnard and S. Hanks, editors, *Uncertainty in Artificial Intelligence 11*, pages 87–98. Morgan Kaufmann, San Francisco, 1995.

[Chickering, 2002a] D.M. Chickering. Learning equivalence classes of Bayesian network structures. *Journal of Machine Learning Research*, 2:445–498, 2002.

[Chickering, 2002b] D.M. Chickering. Optimal structure identification with greedy search. *Journal of Machine Learning Research*, 3:507–554, 2002.

[Meek, 1995] C. Meek. Causal inference and causal explanation with background knowledge. In P. Besnard and S. Hanks, editors, *Uncertainty in Artificial Intelligence 11*, pages 403–410. Morgan Kaufmann, San Francisco, 1995.

[Pearl, 1988] J. Pearl. *Probabilistic Reasoning in Intelligence Systems*. Morgan Kaufmann, San Mateo, CA, 1988.

[Pearl, 2000] J. Pearl. *Causality: Models, Reasoning, and Inference*. Cambridge University Press, NY, 2000.

[Richardson and Spirtes, 2002] T. Richardson and P. Spirtes. Ancestral graph markov models. *Annals of Statistics*, 30(4):962–1030, 2002.

[Spirtes and Richardson, 1996] P. Spirtes and T. Richardson. A polynomial time algorithm for determinint DAG equivalence in the presence of latent variables and selection bias. *Proceedings of the 6th International Workshop on Artificial Intelligence and Statistics*, 1996.

[Spirtes *et al.*, 1993] P. Spirtes, C. Glymour, and R. Scheines. *Causation, Prediction, and Search*. Springer-Verlag, New York, 1993.

[Spirtes *et al.*, 1997] P. Spirtes, T. Richardson, and C. Meek. The dimensionality of mixed ancestral graphs. Technical Report CMU-83-PHIL, Carnegie-Mellon University, Department of Philosophy, Pittsburg, PA, 1997.

[Verma and Pearl, 1990] T. Verma and J. Pearl. Equivalence and synthesis of causal models. In P. Bonissone et al., editor, *Uncertainty in Artificial Intelligence 6*, pages 220–227. Elsevier Science, Cambridge, MA, 1990.